\documentclass[twocolumn,prb,superscriptaddress,showpacs,amsmath,
amssymb,floatfix]{revtex4}
\usepackage{graphicx} \usepackage[english]{babel}

\begin{document}                          

\title {Magnetic collective mode in underdoped cuprates: 
a phenomenological analysis}

\author{P. Prelov\v sek}
\affiliation{Stefan Institute, SI-1000 Ljubljana, Slovenia}
\affiliation{Faculty of Mathematics and Physics, University
of Ljubljana, SI-1000 Ljubljana, Slovenia} 

\author{I. Sega}
\affiliation{Stefan Institute, SI-1000 Ljubljana, Slovenia}

\date{\today}

\begin{abstract}
The dynamical spin susceptibility as relevant for underdoped cuprates is
analysed within the memory-function (MeF) approach. A phenomenological
damping function combined with a $T$-independent sum rule is used to
describe the anomalous normal state and the resonant peak in the
superconducting state, in particular its position and its relative
intensity to the normal state. The relation with the random-phase
approximation is discussed. The MeF method is generalized to the bilayer
system in order to compare with inelastic neutron scattering
experiments on YBa$_2$Cu$_3$O$_{6+x}$ which alows also for a quantitative
comparison. In this context the problem of the missing integrated
spectral intensity within the experimentally accessible energy window is
also discussed.
\end{abstract} 
\maketitle 

\section{Introduction}

Since the first observation of the resonant magnetic peak in
superconducting (SC) optimally doped YBa$_2$Cu$_3$O$_{6+x}$ (YBCO),
\cite{ross} the magnetic collective mode in cuprates and its role  
have been the subject of intensive experimental investigations, with
the most direct information gained by the inelastic neutron scattering
(INS). In YBCO it has been observed that with lower doping the resonant
peak (RP) moves to lower energies, increasing at the same time in
intensity.\cite{bour} In the last years, the intriguing feature of the RP
hour-glass dispersion \cite{arai} has been in the focus. Recently,
also the existence of an optical (even) resonant mode in the bilayer YBCO has been
reconsidered.\cite{pail} It has been shown that the even mode can be well
identified in underdoped (UD) regime, however with an essentially
different intensity from the dominating odd mode. 
On the other hand, it seems clear that the RP has to be closely related to
the magnetic response in the normal (N) state, which has been well resolved
by INS in UD cuprates. The response is typically that of an overdamped
collective mode, however with an anomalous $\omega/T$ scaling, being common
both to UD YBCO \cite{fong} and single-layer La$_{2-x}$Sr$_x$CuO$_4$  
\cite{keim,kast} cuprates.

Most theories, which address the RP and magnetic response in cuprates,
are based on the treatment of the metallic system close to the
antiferromagnetic (AFM) instability, describing the RP as a
consequence of the AFM (over)damped soft mode in the N state and
$d_{x^2-y^2}$ SC gap in the electron-hole excitation spectrum, leading
to a sharp RP below $T_c$. Most frequently invoked is the RPA-like
form for the dynamical susceptibility $\chi_{\mathbf q}(\omega)$,
derived in various ways or argued via the Hubbard, $t$-$J$ or
analogous models. \cite{lava} The  latter 
ascribe the RP to a weak excitonic mode below the electron-hole
continuum and seem to  account qualitatively for optimum doping and for the
overdoped regime, in particular for the position and the peculiar downward
dispersion of the RP. \cite{erem} The alternative approach, using the
memory-function (MeF) description, recently introduced by the present
authors, \cite{spb,psb} gives analogous results in the latter cases,
and in  addition  a more strongly pronounced upper dispersion. \cite{sp}

It is rather evident that the RPA-based theories for the magnetic
response are less appropriate for the UD cuprates, even if taken in a
broader sense as the phenomenological framework. In the first place,
in the N state the usual RPA leads to $T$-independent and
Fermi-liquid-like $\chi_{\mathbf q}(\omega)$, in contrast to the
anomalous dynamics found by INS. \cite{keim,kast} As discussed
furtheron in more detail, the RPA-like description cannot account for
a strong RP in UD cuprates and the spin-wave-like dispersion at higher
energies, also observed in INS experiments. \cite{stoc,bofo} On the other
hand, the MeF approach to $\chi_{\mathbf q} (\omega)$ is able to give
a unified description of the anomalous N-state as well as  SC-state response,
even more naturally in the UD regime. In this paper we will use the
MeF formalism on the phenomenological level with some simplifying
assumptions. It will be shown, that such an approach, generalized to
a bilayer systems relevant for YBCO, can account for the $T$ and
doping evolution of $\chi_{\mathbf q}(\omega)$ in the UD regime,
in particular for the intensity of the coherent RP, and its jump at
the onset of SC in relation with its position.  The framework is also
used to make a quantitative comparison with INS results for UD
YBCO. 

The paper is organized as follows. In the following section we briefly
sketch the MeF formalism (Sec.~II.A) and then address the
evolution  of $\chi''_{\bf q}(\omega)$  with T, including the appearance of
the RP and of its intensity relative to the sum rule $C_{\bf q}$
(Sec.~II.B). Next we present a critical comparison of the MeF approach to
the more commonly used RPA
(Sec.~II.C) and conclude the section with an extension of the MeF approach
to a bilayer system, as appropriate to YBCO. In Section III we first use the
available experimental data on YBCO$_{6.5}$ and YBCO$_{6.7}$ to extract the
relevant parameters within MeF, and then present a quantitative comparison
of the MeF approach to experiments. Conclusions are presented in Sec.~IV.  

\section{Memory function approach}
\subsection{Formalism}

Within the memory function formalism the dynamical spin susceptibility
can be generally written as \cite{spb}
\begin{equation}
\chi_{\mathbf q}(\omega)=\frac{-\eta_{\mathbf q}}{\omega^2+\omega M_{\mathbf
q}(\omega) - \delta_{\mathbf q}} \,, \qquad \delta_{\mathbf q}=
\omega^2_{\mathbf q}
=\frac{\eta_{\mathbf q}}{\chi^0_{\mathbf q} }, \label{eq1}
\end{equation}
where $\chi^0_{\mathbf q}=\chi_{\mathbf q}(\omega=0)$ and the 'spin
stiffness' $\eta_{\mathbf q}=-{\dot\iota}\langle [S^z_{-\mathbf
q}\,\dot{S}^z_{\mathbf q})]\rangle $ can be expressed with equal-time
correlations for any microscopic model of interest and is in general a
quantity only weakly dependent on ${\mathbf q},T$ and even on hole
doping $c_h$.\cite{spb} The latter is not the case for
$\chi^0_{\mathbf q}$ or for the 'mode' frequency $\omega_{\mathbf  q}$.
Instead of searching for an explicit approximation for either of these
quantities, we fix them via the fluctuation-dissipation sum rule,
\cite{spb,psb} 
\begin{equation}
\frac{1}{\pi}\int_0^\infty d\omega ~{\rm cth}\frac{\omega}{2T}
\chi^{\prime\prime}_{\mathbf q}(\omega)= \langle S^z_{-{\mathbf q}} 
S^z_{\mathbf q}\rangle = C_{\mathbf q} . \label{eq2}
\end{equation}
The underlying idea is that the equal-time spin correlations $C_{\mathbf q}$
are much less sensitive on $T$. In particular, we conjecture that they
do not change (significantly) at the N-SC transition, which we lateron
verify for available YBCO data. Evidently, $C_{\mathbf q}$ is ${\mathbf q}$ and
doping dependent, but note that in strongly correlated systems $C_{\mathbf
q}$ is restricted by the total sum rule, $(1/N)\sum_{\mathbf q} C_{\mathbf q}
=(1-c_h)/4$. In the following, we concentrate in the analysis to the
commensurate ${\mathbf Q}=(\pi,\pi)$, so the relevant quantity is
$C_{\mathbf Q}$.  

Here, we do not intend to give a microscopic derivation for the
damping function $\gamma_{\mathbf q}(\omega)=M^{\prime\prime}_{\mathbf
q}(\omega)$, as performed within the $t$-$J$ model. \cite{spb,psb,sp} Still,
we use the observation that the low-$\omega$ 
damping within the N state of a doped AFM, being metallic and
paramagnetic, is mainly due to the electron-hole excitations. If the Fermi
surface of the doped system crosses the AFM zone boundary (as revealed
by ARPES for most hole-doped cuprates), in the N state one gets
$\gamma_{\mathbf q}(\omega \to 0) >0$ for ${\mathbf q}\sim {\mathbf
Q}$. Hence, we use as the phenomenological input in the N state the
simplest approximation $\gamma_{\mathbf q}(\omega)=\gamma$, supported
also by numerical calculations within the $t$-$J$ model. \cite{psb}.
The expression for $\chi_{\bf q}(\omega)$, Eq.(\ref{eq1}), in the N state
then reduces to a simple damped-oscillator form. 

In the SC state the d-wave gap introduces a gap also into
$\gamma_{\mathbf q}(\omega)$. Here, we are not interested in the
dispersion of the RP, \cite{erem,sp} but rather in particular
${\mathbf q}={\mathbf Q}$. Hence, we assume at $T<T_c$
\begin{equation}
\gamma_{\mathbf Q}(\omega<\omega_c)=0, \qquad
\gamma_{\mathbf Q}(\omega>\omega_c)=\gamma,
\end{equation}
where the effective gap $\omega_c=2\Delta({\mathbf q^*}) <\Delta_0 $ is
given by the SC gap value at ${\mathbf q^*}$ where the FS crosses the AFM
zone boundary. For the discussion of the RP the corresponding real part is
essential, 
\begin{equation}
M^\prime_{\mathbf Q}(\omega) =\frac{\gamma}{\pi} \ln
|\frac{\omega_c+\omega}{\omega_c-\omega} |, \label{eq3}
\end{equation}
generating always an undamped excitonic-like RP at $\omega_r<\omega_c$.

\subsection{Qualitative analysis within the MeF}

The presented phenomenological theory has nontrivial consequences and
predictions for the spin dynamics in cuprates, both for the N and thr SC
state.  It follows directly from Eq.(\ref{eq1}) that the
damped-oscillator form is well adapted to treat a collective mode in the 
UD regime close to the AFM instability. Here we focus on the 
${\mathbf q}={\mathbf Q}$ mode and take into account also the fact, that in
the N state the latter is generally overdamped, requiring $\gamma> 
\omega_{\mathbf Q}$. Then, we get essentially two rather distinct regimes:
a) if $\omega_{\mathbf Q} \gg \omega_c$, consistent with modest $C_{\mathbf
Q} < 1$, the N-state $\chi^{\prime\prime}_{\mathbf Q}(\omega)$ is very
broad leading to a weak excitonic-like RP at $\omega_r \lesssim
\omega_c$. This is the situation corresponding to optimum-doped or
overdoped cuprates. b) When $\omega_{\mathbf Q}$ is close or even
below $\omega_c$, requiring $C_{\mathbf Q} > 1$, there is a pronounced
low-$\omega$ response already in the N state, transforming into a
strong RP in the SC phase, exhausting a substantial part of the sum
rule $C_{\mathbf Q}$.

In Fig.~1 we present typical $\chi^{\prime\prime}_{\mathbf
Q}(\omega)$, corresponding to the UD regime. We fix the parameters from
our  knowledge of the results within the $t$-$J$ model, \cite{spb} as
relevant for cuprates. For a weakly doped AFM we adopt $\eta \sim 2J
\sim 0.6t$ (note that $t\sim 400$ meV), the value for an
undoped AFM. In the UD regime it is essential that the damping is
small, $\gamma < t$, both to get reasonable low-$\omega$ response as
well as underdamped spin waves at higher $\omega$. Here, we assume
$\gamma =0.2 t$, while the SC parameters are chosen in accordance with
data for UD cuprates, e.g., $\omega_c=0.15~t$, $T_c = 0.05~t$. Then
$C_{\mathbf Q}$ enters into $\chi^{\prime\prime}_{\mathbf Q}(\omega)$
as the only ''free'' parameter. In spite of simplifications, we
observe in Fig.~1(a) several features consistent with experiments: a)
the mode is overdamped in the N state with a Lorentzian form
$\chi^{\prime\prime}_{\mathbf Q}(\omega) \propto
\omega/(\omega^2+\Gamma^2)$ and a peak $\Gamma \propto T$ shifting
with $T$, being the signature of the anomalous $\omega/T$ scaling,
observed in UD cuprates. \cite{keim,kast} b) Already at $T>T_c$ the
peak $\Gamma$ is below $\omega_c$, and in fact close to the position
of the RP, which is not just a coincidence.  c) At $T<T_c$ there
appears a strong RP, exhausting a large part of the sum rule
$C_{\mathbf Q}$. Also it is well shifted from the effective gap, i.e.,
$\Delta \omega_r= \omega_c - \omega_r$ is not small. 
\begin{figure}
\begin{center}\includegraphics[%
    width=60mm,angle=-90,
    keepaspectratio]{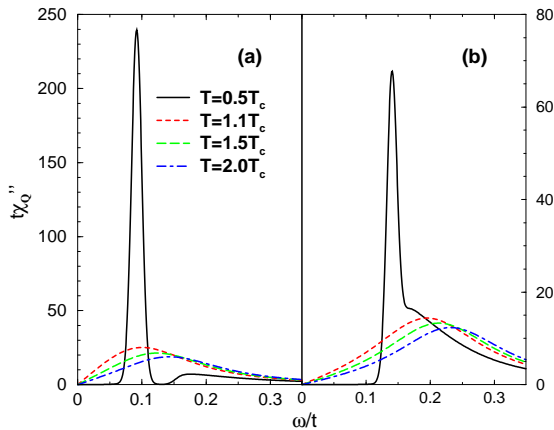}\end{center}                            
\caption{\label{cap:Fig1}(Color online) Dynamical spin
susceptibilities $\chi^{\prime\prime}_{\mathbf Q}(\omega)$ for
different $T$ above and below $T_c$, with other parameters fixed as
given in the text. (a) $C_{\mathbf Q}=2$, and (b)
$C_{\mathbf Q}=1$. Spectra are additionally broadened with $\delta =
0.01~t$. }
\end{figure}
The presented case is very close to the actual INS results for YBCO at
$x=0.5- 0.7$, \cite{bour,fong,stoc} as discussed lateron.  

We define the intensities as 
\begin{equation}
I_{\mathbf Q} =\int d\omega \chi^{\prime\prime}_{\mathbf Q}(\omega).
\end{equation}
In Fig.~2 we follow the RP intensity $I_r$, as a function of the
relative RP position $\Delta \omega_r/\omega_c$, where we vary
$C_{\mathbf Q}$ and fix the other parameters. In this way, we cover the
span between the weak excitonic RP (small $C_{\mathbf Q}$) up to a
pronounced AFM soft mode (large $C_{\mathbf Q}$).  In INS experiments
the strength of the RP is usually presented as the difference between
the SC- and N-state response, \cite{fong} i.e., as 
\begin{equation}
\Delta I_r=I_r(T \sim 0)-I_r(T\agt T_c),
\end{equation}
hence we plot in Fig.~2 also $\Delta I_r$, with subtracted N-state intensity
at $T>T_c$ and $\chi^{\prime\prime}_{\mathbf Q}(\omega)$ integrated in the
range $0< \omega<\omega_c$. As discussed in Sec.~IID, such a measure can be
directly compared with the recent INS experiments. \cite{pail} As noticed
elsewhere, \cite{mill,pail} it follows from the singular behavior of
$M'({\mathbf Q},\omega)$, Eq.~(\ref{eq3}), that $I_r \propto \Delta \omega_r$,
valid in the regime of weak RP, $\Delta \omega_r \ll \omega_c$. From
Fig.~2 one infers that a nearly linear dependence remains valid well
beyond this limit. Also the RP enhancement $\Delta I_r$ shows similar
dependence although overall reduced. Note that in the extreme case
$\Delta I_r$ can become even slightly negative. It is also worth
observing that $\Delta I_r$ vs. $\Delta \omega_r$ is only weakly
dependent on $\gamma$, which might explain why YBCO with different
doping $x$ can be plotted roughly on a unique curve.\cite{pail}
\begin{figure}
\begin{center}\includegraphics[%
    width=60mm,angle=-90,
    keepaspectratio]{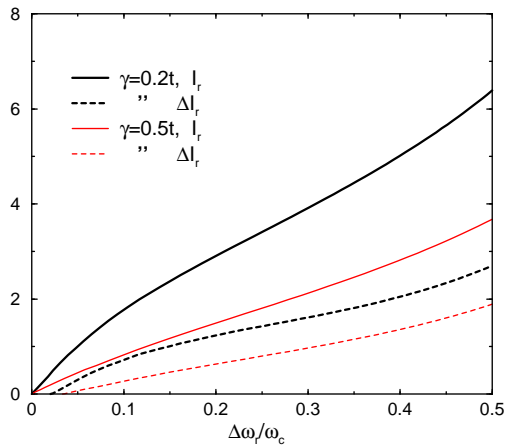}\end{center}                            
\caption{\label{cap:Fig2}(Color online) Resonant peak intensities
$I_r$ (full lines) and $\Delta I_r$ (dashed lines), without and with
N-state intensity subtracted, vs. relative position $\Delta \omega_r/\omega_c$,
for two values of $\gamma$ and varying $C_{\mathbf Q}$.}
\end{figure}

\subsection{Comparison with RPA}

Let us also comment on the similarities and differences with the more usual 
RPA-like representation
\begin{equation}
\chi_{\mathbf q}(\omega)= [ \tilde \chi_{\mathbf q}(\omega)^{-1} - 
\tilde J_{\mathbf q}]^{-1}, \label{eq4}
\end{equation}
where in the SC state at $T=0$ 
\begin{eqnarray}
\tilde \chi_{\mathbf q}(\omega)&=&-\frac{1}{N} \sum_{\mathbf k} \Bigl
(1-\frac{\tilde \epsilon_{{\mathbf k}+{\mathbf q} } \tilde
\epsilon_{\mathbf k} + \Delta_{{\mathbf k}+{\mathbf q}}
\Delta_{\mathbf k} }{ E_{{\mathbf k}+{\mathbf q}} E_{\mathbf k} }
\Bigr ) \nonumber\\ &\times &\frac{ E_{{\mathbf k}+{\mathbf q}} +
E_{\mathbf k} } { \omega^2- (E_{{\mathbf k}+{\mathbf q}} + E_{\mathbf
k} )^2 }, \label{eq5}
\end{eqnarray}
with $E_{{\mathbf k}}=\sqrt{{\tilde \epsilon}_{\mathbf k}^{\,2}+
\Delta_{\mathbf k}^2}$, whereas in the N state $\tilde \chi_{\mathbf q}$
corresponds to the usual Lindhard function, and 
%$\tilde\epsilon_{\mathbf k}= -4\tilde t \tilde \gamma_{\mathbf k} -4\tilde
%t^\prime \tilde \gamma^\prime_{\mathbf k}-\mu $ 
$\tilde\epsilon_{\mathbf k}=-2\tilde t [\cos(k_xa)+\cos(k_ya)]-4\tilde
t^\prime\cos(k_xa)\cos(k_ya)-\mu$ is the quasiparticle (QP)
dispersion.  The form, Eq.(\ref{eq4}), has been derived in several
ways or postulated based on microscopic models such as the Hubbard
model and the $t$-$J$ model. \cite{lava,erem} We treat
it as a phenomenological expression where parameters $\tilde
J_{\mathbf Q},\tilde t,\tilde t'$ are $T$ independent, while still
possibly doping dependent.

In spite of the apparent different forms, RPA and MeF, Eq.~(\ref{eq1}), can
be related at low $\omega$. If the QP band crosses the AFM zone boundary,
we get in the N state $\tilde \chi^{\prime\prime}_{\mathbf Q}(\omega \to
0) = \tilde \Gamma_{\mathbf Q} \omega$. Using the relation
\begin{equation}
( \tilde \chi_{\mathbf Q}(\omega)^{-1}-\tilde J_{\mathbf Q})\eta_{\mathbf Q}
= \delta_{\mathbf Q}-\omega^2-\omega M_{\mathbf Q}(\omega), \label{eq6}
\end{equation}
it follows
\begin{equation}
\gamma=M^{\prime\prime}_{\mathbf Q}(\omega\to 0)=
\eta \tilde \Gamma_{\mathbf Q}/
(\tilde \chi^0_{\mathbf Q})^2 , 
\label{grpa}
\end{equation}
while $\eta_{\mathbf Q}=-[\omega^2 \tilde \chi_{\mathbf Q}]_{\omega \to
\infty}\propto \tilde t$. Since in the SC state $\tilde \chi^{\prime\prime}_{\mathbf
Q}(\omega<\omega_c)=0$, the RPA result is also at $T<T_c$ formally
close to our phenomenological MeF approach.

Nevertheless, the qualitative and even more quantitative differences
become very pronounced in the UD regime approaching the undoped AFM:
a) in the N state $\chi^{\prime\prime}_{\mathbf q}(\omega)$,
Eq.~(\ref{eq4}), is essentially $T$ independent and cannot account for
the N-state anomalous scaling. b) For all reasonable QP bands,
the effective $\gamma$ within the RPA obtained from Eq.~(\ref{grpa}) is very
large, i.e., typically $\gamma>t$. This prevents underdamped spin
waves even at large $\omega$. c) The intensities $I_r$ of the RP  are
generally small, and in particular the N-SC difference $\Delta I_r \sim 0$. 

For illustration, we present in Fig.~3  characteristic results within the RPA
approach in the N and SC state $T \sim 0$, which correspond as close
as possible to the regime of MeF results in Fig.~1. We choose the
same $\omega_c$, while $\tilde t=3\tilde t'=0.3t$. Data in Fig.~3(a)
correspond to $\tilde J_{\mathbf Q} = 0.95 \tilde J_{\mathbf Q}^{c}$,
i.e., close to the AFM instability. Although the RP is quite intense
in this case, exhausting $~60\%$ of the sum rule, the respective
$I_r\sim1.0$ is considerably smaller in comparison to both the MeF
case (Figs.~1(a),2) and the INS data given later (see
Table~\ref{table1}). On entering the SC state $C_{\mathbf Q}$ exhibits
a drop in magnitude, which may be quite substantial for $\tilde
J_{\mathbf Q}$ close to $\tilde J_{\mathbf Q}^c$, again in contrast
with experiments. Similarly, for $\tilde J_{\mathbf Q} = 0.75 \tilde
J_{\mathbf Q}^{c}$ (Fig.~3(b)) where the RP intensity still accounts for
$\sim 20\%$ of the sum rule, we get $I_r\sim 0.25$, one order of magnitude
smaller than the experimental data.  Also note that $\Delta I_r$ remains
negligeable ($\alt 0.3$) for all $\tilde J_{\mathbf Q}<\tilde
J_{\mathbf Q}^c$.
\begin{figure}
\begin{center}\includegraphics[%
    width=60mm,angle=-90,
    keepaspectratio]{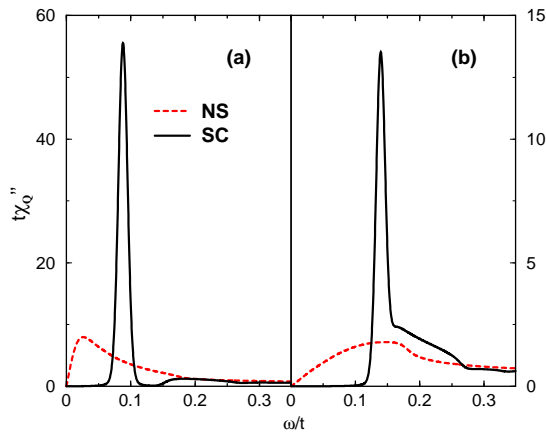}\end{center}
\caption{\label{cap:Fig3}(Color online)$\chi_{\mathbf
Q}^{\prime\prime}(\omega)$ as evaluated within the RPA in the N state 
and SC state, respectively. (a) $\tilde J_{\mathbf Q}=0.95\tilde J_{\mathbf
Q}^c$, (b) $\tilde J_{\mathbf Q}=0.75\tilde J_{\mathbf Q}^c$.}   
\end{figure}

\subsection{Coupled-layers analysis}

In order to be able to describe more quantitatively the INS results
for YBCO, which is a bilayer system, we generalize the MeF approach
to two coupled layers. We assume that the interaction between layers
is only via an isotropic exchange \cite{mill}
\begin{equation}
H_{12}=J_\perp\sum_i{\mathbf S}_{i1} \cdot  {\mathbf S}_{i2}.
\end{equation}
Susceptibility can be defined as a matrix $\chi^{ll^\prime}_{\mathbf
q}(\omega)$, as well as other quantities, $\delta^{ll'}_{\mathbf
q},\eta^{ll'}_{\mathbf q}$ etc. The plausible assumption we use here
is that the dominant damping due to the electron-hole excitations can
originate only from the intraplanar hopping, therefore
$M^{12}=0$. Taking into account the symmetry $\chi^{11}=\chi^{22}$
etc. and defining even and odd functions, respectively,
$\chi^{e,o}_{\mathbf q}=\chi^{11}_{\mathbf q} \pm \chi^{12}_{\mathbf
q}$ etc., we can write in analogy with Eq.~(\ref{eq1})
\begin{equation}
\chi^{e,o}_{\mathbf q}=\frac{-\eta^{e,o}_{\mathbf q} }
{ \omega^2+\omega M_{\mathbf q}(\omega) - \delta^{e,o}_{\mathbf q} },
\label{eq11}
\end{equation}
where $\delta^{e,o}_{\mathbf q}$ are related $C^{e,o}_{\mathbf q}$ via the
sum rules as before ( Eq.(\ref{eq2})). Also, $\eta^{e,o}_{\mathbf q}$ can be
explicitly expressed for the above interplane coupling $H_{12}$, and
we get 
\begin{equation}
\Delta \eta =\eta^{o}_{\mathbf Q}-\eta^{e}_{\mathbf Q} \propto
J_\perp \langle {\mathbf S}_{1i} \cdot {\mathbf S}_{2i}\rangle 
< (J_\perp/J) \eta.
\end{equation}

The estimates in the literature are $J_\perp/J \sim 0.1$, \cite{hayd}
therefore to the lowest approximation we can neglect the difference,
and $\eta^{e,o}_{\mathbf Q} \sim \eta$. The only appreciable source of
distinction between the even and odd response is therefore
$C^o_{\mathbf Q}>C^e_{\mathbf Q}$. The latter difference can become
substantial in UD cuprates which are close to the AFM instability,
where the soft mode is $\omega^o_{\mathbf Q}$ while $\omega^e_{\mathbf
Q} \propto \sqrt{J_{\perp}J}$ remains gapped. \cite{hayd}

With the above simplifications we can apply previous single-plane
results directly to bilayer YBCO. In particular, in Fig.~2 results for
$I_r, \Delta I_r$ vs. $\Delta \omega_r$ for even and odd mode should
fall on the same quasi-linear curve (for same $\gamma, \eta$),
corresponding to different $C_{\mathbf Q}$.  E.g., we note that
$\Delta \omega_r/\omega_c = 0.05, 0.3$ correspond (for $\gamma=0.2~t$)
to $C_{\mathbf Q}=1.2, 1.75$, respectively, being qualitatively
consistent with INS experiments. \cite{fong}

\section{M$\rm e$F approach $\rm vs$ experiments}

Let us use the existing quantitative INS data for YBCO to check
the relevance of the presented MeF analysis. First, the spectra for
YBCO with $x=0.5$ \cite{stoc} can be used to extract directly
$\gamma$.  Namely, at higher $\omega \sim 80$~meV the spin-wave branches
become underdamped and well resolved. From the width we can then
estimate $\gamma \sim 70$~meV~$\sim 0.2 t$, the value assumed in our
presentation in Figs.~1(a),2(a). We notice that such low $\gamma$ (as well as
spin waves) is essentially impossible within the RPA approach.

Next, we extract data for the intensities $I_{\mathbf Q}$ and the sum
rule $C_{\mathbf Q}$, Eq.~(\ref{eq2}), for UD YBCO, both in the N and
the SC state. To our knowledge such results are only available for $x=0.7$
\cite{fong} and $x=0.5$. \cite{bour} Clearly, the spectra we can only
integrate in the measured window $\omega <70$~meV, so results the 
represent a lower estimate. 
\begin{table}
\begin{ruledtabular}
\caption[TABLE I] {Values extracted from INS results for UD YBCO (odd
mode). $C_{\mathbf Q}$ and $I^{\mathrm exp}(=4I\mu_B^2$), are estimated
from the data published in $^{a)}$~Ref.~\onlinecite{bour},
$^{b)}$~Ref.~\onlinecite{stoc}, and $^{c)}$~Ref.~\onlinecite{fong}.}
\label{table1}
\begin{tabular}{c|ccc|ccc}                     
& & YBCO$_{\,6.5}$ & & & YBCO$_{\,6.7}$\footnotemark[3] & \\ \colrule
T [K] & 5\footnotemark[1] & 85\footnotemark[2] & 100\footnotemark[1] &
12 & 70 & 200 \\ $C_{\mathbf Q}$[1/f.u.] &2.1 &1.6 & 1.9 &1.4 & 1.4 &
1.5 \\ $I_{\mathbf Q}^{\mathrm exp}[\mu_B^2$/f.u.] & 26.4 & 18.0 &
11.3 & 16.9 &14.9 & 8.8 \\ $I_r^{\mathrm exp}[\mu_B^2$/f.u.] &
19\footnotemark[2]& & & 12.6 & & \\
\end{tabular}       
\end{ruledtabular}
\end{table}                             

In Table~\ref{table1} we present data for the odd mode, where the
low-$\omega$ contribution is dominant. The important message is that
$C_{\mathbf Q}$ is indeed quite $T$-independent and in particular nearly
conserved at the N-SC transition, being the essential assumption within our
MeF approach. Also,  (theoretical) intensities $I_{\mathbf Q}^o, I_r^o$ are
both large and comparable with MeF results in Fig.~2. 

Finally, we present an attempt of a quantitative fit of the same INS
data \cite{bour,fong} using our phenomenological description as
presented in Sec.~IIA. We omit the YBCO$_{6.7}$ data at $T=85$ K,
which is just above $T_c$ as there seems to be already an indication
for a reduced N-state damping, since the response appears already
underdamped. In adjusting the MeF form to the INS data we adopt the same 
form for $M_{\bf Q}(\omega)$ as in Sec.~IIA, but for $T<T_c$ we keep
$\gamma_{\mathbf Q}(\omega<\omega_c)=\gamma_0$ finite since experimentally
the RP for $T$ well below $T_c$ seems not to be resolution limited,i.e., the
RP has a finite width. Thus: 
\begin{eqnarray}
M''_{\bf Q}(\omega)&=&\gamma_0 \theta(\omega_c-\omega)+\gamma \theta
(\omega-\omega_c)\,,\\ M'_{\bf
Q}(\omega)&=&\frac{\gamma-\gamma_0}{\pi}\ln\vert
\frac{\omega_c+\omega}{\omega_c- \omega}\vert\,,
\label{eqd}
\end{eqnarray}
where $\theta(x)$ is the step function. In adjusting the MeF
form we try to keep $C_{\bf Q},\eta=\eta_{\bf Q}, \gamma$ as
$T$-independent as possible. As before, we fix the effective SC gap at
$\omega_c\sim 60$ meV, consistent with experimental analysis
\cite{pail}. The resulting parameters are listed in Table~II, while
the fits are presented in Fig.~4. 
\begin{figure}
\begin{center}\includegraphics[%
    width=80mm,angle=0,
    keepaspectratio]{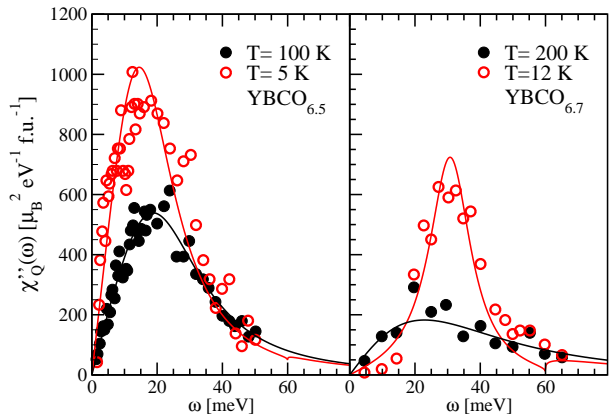}\end{center}                            
\caption{\label{cap:Fig4}(Color online) MeF fit to INS data for
$\chi''_{\bf Q}(\omega)$ (in absolute units) as given in
Refs.\onlinecite{bour,fong} Parameters are given in Table~II.}
\end{figure}
\begin{table}
\begin{ruledtabular}
\caption[TABLE II] {Values adopted in fitting INS data for UD YBCO (odd
mode) as presented in Fig.~4.}
\label{table2}
\begin{tabular}{c|ccc|ccc}                     
& & YBCO$_{\,6.5}$ & & & YBCO$_{\,6.7}$ & \\ 
\colrule
T [K] &5 & & 100 &12 &  & 200 \\
$\eta_{\mathbf Q}$ [meV] &130 & & 120 &130 &  & 120 \\
$\gamma$ [meV] & 35 & & 40 & 60 & & 85 \\
$\omega_{\mathbf Q}$[meV] &20 & & 26 & 38  &  & 40 \\
$\gamma_c$ [meV] &30 & &  &23 &  &  \\
$C_{\mathbf Q}$[1/f.u.] & 2.1 &  & 2.1 & 1.2 & & 1.4 \\
\end{tabular}       
\end{ruledtabular}
\end{table}                

Overall, the fits appear quite satisfactory provided, however, that
one chooses $\eta\sim 120-130$ meV $\sim J$. Essentially, $\eta$
normalizes the intensities. While theoretically (from model
calculations) $\eta_{th}\sim 2J$ is a very robust quantity, there are
several reasons why $\eta_{th}$ becomes renormalized. We are not
dealing with the whole energy spectrum of spin fluctuations, but only
with the low-frequency $\omega<J$ part. It is plausible, although
theoretically not well explored, that there is a substantial
high-frequency $\omega> J$ dynamics, in particular increasing with
doping. 
This is clearly inferred from spectra obtained in numerical
simulations on finite clusters for the $t$-$J$ model at finite temperature,
\cite{jjpp} where at low $T$ and low to moderate doping a rather distinctive
separation of energy scales for $\omega\alt J$ and $\omega\agt 2\tilde t$
occurs. An enhancement of  the sum rule $C_{\bf Q}$ with respect to the
low-$\omega$ region should then occur. But $\eta$ should become enhanced
even more since, being the first frequency moment of  $\chi''_{\bf
Q}(\omega)$, it is more susceptible to high-$\omega$ tails in 
$\chi''_{\bf Q}(\omega)$.
Likewise, a partial reduction of the effective
fluctuating spin would also result in a smaller $\eta$. In any case, the
same question already seems to be present when interpreting the INS results
for undoped AFM where a reduced magnon intensity is observed through a
renormalization factor $Z_\chi\sim 0.5$.\cite{bour1}.

Formally, a renormalization effect can be easily incorporated into the MeF
analysis by adding an effective damping above some threshold
$\omega^*>J$, assuming $M''_{\bf Q}(\omega>\omega^*) \sim \gamma^* \gg
\gamma$ and
\begin{equation}
\Delta M^\prime_{\mathbf Q}(\omega) =\frac{\Delta\gamma}{\pi} \ln \Bigl|
\frac{\omega^*+\omega}{\omega^*-\omega} \Bigr|
\sim \frac{2\Delta\gamma}{\pi \omega^*} \omega\,. 
\end{equation} 
where $\Delta\gamma=\gamma^*-\gamma$. Such a term would renormalize
$\eta$ for $\omega \ll \omega^*$
\begin{equation}
\eta \to \tilde \eta = \eta [1 + \frac{2\Delta\gamma}{\pi\omega^*}]^{-1}\,,
\end{equation}
which could explain the experimentally determined $Z_{\chi}$. Note, however,
that a separation of energy scales as mentioned above would also lead to a
similar result, since a suppression of $\chi''_{\mathbf q}(\omega)$ in the
energy interval $\omega^*<\omega< 2\tilde t$ is directly related to a  local
stepwise enhancement of damping within the same energy interval, i.e.,
$M''_{\mathbf q}(\omega^*<\omega<2\tilde t)\sim\Delta\gamma$.

\section{Conclusions}

To summarize, we have presented a phenomenological analysis of the
magnetic collective mode in UD cuprates within the MeF approach and
have compared it with frequently used RPA in this context.  Both
approaches give some qualitatitvely similar results for the RP, in
particular $I_r \propto \Delta \omega_r$. \cite{mill,pail} However,
there are also clear differences.  Due to large damping $\gamma\sim t$, the
RPA cannot capture the high-energy magnons, \cite{stoc,bofo} 
while these are easily reproduced  within the MeF approach. Within RPA the
RP intensities are too small in comparison to experiments, while $C_{\mathbf
Q}$ drops at the N-SC transition, in contrast with experiments as analysed in 
Table~\ref{table1}. 

On the other hand, MeF shows in the N state an anomalous $\omega/T$
scaling of $\chi_{\mathbf Q}^{\prime\prime}(\omega)$, only being
interrupted by the onset of SC, whereby the RP intensities $I_r$ close
to experimental (see Table~\ref{table1}) can be easily
reproduced. However, there is significant discrepancy between $\Delta
I_r$ and $\Delta I_r^{\textrm exp}/4\mu_B^2$ as the latter is quite
small. In fact, in underdoped cuprates $\Delta I_r^{\textrm exp}$
might not be as relevant a quantity for it compares the RP
(integrated) intensity below $T_c$ with the N-state intensity just
above $T_c$, where the appearance of a pseudogap can already reduce
damping $\gamma$ and induce a RP-like response. \cite{fong,stoc,dai}

Generalizing the MeF approach to a bilayer system as appropriate for
YBCO we have shown that the RP intensities in both channels, i.e.,
$I_r^{e,o}$ should fall on the same quasi-linear curve
vs. $\Delta\omega_r$, in agreement with experiments. \cite{pail}

Finally, we have performed a quantitative fit within the
MeF approach, which agrees well with the INS data, provided that
$\eta$ at low $\omega$ is considerably renormalized. 
\vskip 12pt
The authors  acknowledge Y. Sidis and P. Bourges for the enlightening
discussion. This work was supported by the Slovenian Research Agency under
Grant P-106-017.

\end{document}